\documentclass[%
 reprint,
 amsmath,amssymb
,aps
]{revtex4-1}

\usepackage{graphicx}
\usepackage{dcolumn}
\usepackage{bm}
\usepackage{tabularx, booktabs}
\usepackage{braket}
\usepackage{wrapfig}
\usepackage[colorlinks=true]{hyperref}
\usepackage{color}

\renewcommand{\vr}{\textbf{r}}

\newcommand{\cG}{{\cal G}}
\newcommand{\cU}{{\cal U}}
\newcommand{\Tr}{{\mathrm{Tr}}}

\begin{document}

\preprint{APS/123-QED}

\title{Dynamical Mean Field Theory for Diatomic Molecules and the Exact Double Counting}

\author{Juho Lee}
\author{Kristjan Haule}%
\affiliation{%
Department of Physics \& Astronomy, Rutgers University, Piscataway, NJ 08854-8019, USA
}%

\date{\today}

\begin{abstract}
Dynamical mean field theory (DMFT) combined with the local density approximation (LDA) is widely used in solids to predict properties of correlated systems. In this study, a parameter-free version of LDA+DMFT framework is implemented and tested on one of the simplest strongly correlated systems,  H$_2$ molecule. Specifically, we propose a method to calculate the exact intersection of LDA and DMFT that leads to highly accurate subtraction of the doubly counted correlation in both methods. When the exact double-counting treatment and a good projector to the correlated subspace are used, LDA+DMFT yields very accurate total energy and excitation spectrum of H$_2$. We also discuss how this double-counting scheme can be extended to solid state calculations.
\end{abstract}

\maketitle



\section{Introduction}

Quantum mechanics has long sought deeper insight into correlation
effects because they lie at the heart of understanding atomic,
molecular and solid state electronic structure. 
Over the past few decades, many theoretical frameworks have been developed
to describe so-called ``strongly correlated systems", in which quasi-particle
approaches such as density functional theory (DFT) \cite{kohn-hohenberg1964,kohn-sham1965}
essentially fail due to the large Coulomb interaction between electrons.
Among them, dynamical mean field theory (DMFT) \cite{georges-kotliar1992,kotliar1996rev} has 
brought about a revolution in the theory of strong
correlations after its exact treatment of local dynamic
correlations successfully described the Mott transition in lattice
models such as the Hubbard model \cite{kotliar_physicstoday}. Since
the method is very flexible and versatile, and scales linearly with
the system size, it has been quickly adapted for many solid state problem, 
including electronic structure calculations.

The most commonly used DMFT approximation in solid state is combination of LDA
and DMFT (LDA+DMFT) \cite{kotliar2006rev}, where some selected
correlated orbitals are treated by DMFT while the rest of the
electronic states are treated by LDA. The LDA+DMFT method has been very
successful in various problems involving strong electronic
correlations in solids and very recently it was also applied to
molecules~\cite{millis2011chem,zgid2011} and
nano-systems \cite{haule_molecule,dmft_nano2012}. However,
the application of this methodology to solids has a few
ambiguities, which limit the precision of the method: i) the DMFT method
needs the partially screened Coulomb interaction, which is hard to
predict from first principles, ii) the part of the
correlations treated by both LDA and DMFT -- called double-counting
-- is not known exactly, and a phenomenological form~\cite{dc_fll}
has been most often used (for discussion see \cite{dc_haule} and
\cite{dc_millis}).

The study of correlations in small molecules can be a testbed for
the quality of electronic structure methods because numerically exact
results exist. From the DMFT view point, this has a particularly strong appeal
because the screening of the Coulomb repulsion can be negligible and
therefore the ambiguities due to screening, present in solid state, can be
decoupled from the issues concerning construction of the
functional and its precision. In addition, the short-range nature of 
dynamical correlation in molecules  \cite{mol_loc_corr1,mol_loc_corr2}
further justifies the applicability of DMFT to molecules.
While H$_n$ clusters \cite{millis2011chem} and H- cubic solid \cite{zgid2011} 
have been investigated within the DMFT framework,
the simplest case of H$_2$ molecule,
which shows very strong correlations at large nuclear separation, 
has not been  studied yet by DMFT.

In this paper we propose a
double-counting functional for LDA+DMFT, which is an exact
intersection of the two methods and results in highly precise
electronic structure method with no ambiguity in subtracting
doubly counted correlation effect. We also suggest the extension of this
double-counting functional to the solid state calculations, where
additional complexity of screening will need to be addressed.
Our basis set is the eigenfunctions of H$_2^+$ 
exactly solved by the methodology of
Ref.~\onlinecite{hadinger1989}. We denote the ground state and the first
excited state by $\ket{1\sigma_g}$ and $\ket{1\sigma_u}$,
respectively. Typically between other 20-30 excited states are used as a basis for H$_2$
calculation for good convergence.

Since DMFT is a basis set dependent approximation, 
its quality depends essentially on the choice of the projector \cite{kotliar_local,haule_local,Vanderbilt_WF,pavarini_WF,anisimov_WF,amadon_WF}, 
which maps the continuous
problem to a discrete set of sites (lattice), each consisting of only
a few important degrees of freedom (orbitals).
In this work, we restrict our discussion to the simplest possible DMFT
approximation, taking only one correlated orbital per site.
Since the two sites are equivalent by the symmetry, 
the problem reduces to a single site one orbital impurity problem, which can be solved
to very high precision by the continuous time quantum Monte Carlo
method~\cite{Werner2006}, as implemented in
Ref.~\onlinecite{haule_ctqmc}.

\section{Theory and Method}

A good choice of the DMFT projector should have large overlap with the most active states around the Fermi level, should be well localized on an atom, recover atomic solution in the large separation limit, and finally should not depend on the self-consistent charge density. Without the last condition, 
it is impossible to obtain a stationary solution by extremizing Luttinger-Ward functional.

A natural choice of the DMFT
projector of this H$_2$ problem is the linear combination of 
the lowest bonding $\ket{1\sigma_g}$ and anti-bonding state $\ket{1\sigma_u}$ of H$_2^+$ system, which we
define as the ``left" (L) and the ``right" (R) localized orbital,
\begin{align}\label{loc_orbital}
|\chi_L \rangle = \frac{1}{\sqrt{2}} ( |1\sigma_{g}\rangle - |1\sigma_{u}\rangle),\nonumber \\
|\chi_R \rangle = \frac{1}{\sqrt{2}} ( |1\sigma_{g}\rangle + |1\sigma_{u}\rangle).
\end{align}
that naturally recover 1$s$ state of each site at large atomic separation.
Over 96\% of the electronic charge of the DMFT solution is contained
in these two states and since they do not explicitly depend on the DMFT Green's function,
these are a good choice for DMFT orbital.

We define the DMFT local Green's function for left atom by the
projection $\cG^{L}_{local}(\omega)\equiv \hat{P}_L
G=\ket{\chi_L}\bra{\chi_L}G(\vr,\vr',\omega)\ket{\chi_L}\bra{\chi_L}$
and similarly for the right atom. The impurity self-energy is embedded
into real space by the inverse of the projection, i.e.,
$\Sigma(\vr,\vr',\omega)=\ket{\chi_L}\Sigma^L(\omega)\bra{\chi_L} +
\ket{\chi_R}\Sigma^R(\omega)\bra{\chi_R}$.  Due to the symmetry of the
problem, $\Sigma^L=\Sigma^R$ and $G^L=G^R$. We mention in passing that
the alternative choice of projector, which selects as the correlated
orbital $1s$ state of each atom, leads to a result of 
worse quality than presented here, because such choice does not
capture the majority of the active degrees of freedom at equilibrium
internuclear separation. Consequently, more time consuming
cluster-DMFT method needs to be used to obtain similar quality results,
as recently found in Ref.~\onlinecite{millis2011chem}.

To construct the DMFT framework, we resort to the Baym-Kadanoff formalism
\cite{baym_kadanoff,baym1962}, which defines a functional of the full Green's
function by (see also \cite{kotliar2006rev})
\begin{equation}\label{bk}
\Gamma[G] = \Tr\log(G)-\Tr((G^{-1}_{0}-G^{-1})G)+\Phi[G].
\end{equation}
Here $G_{0}(\mathbf{r},\mathbf{r}';i\omega) =[(i\omega + \mu +\nabla^2 -V_{ext}(\mathbf{r}))\delta(\mathbf{r}-\mathbf{r}')]^{-1}$,
and $V_{ext}$ is the potential created by the two nucleus.
$\Phi[G]$ is the so-called Luttinger-Ward functional 
and is equal to the sum of all skeleton Feynman diagrams 
consisting of $G$ and Coulomb interaction $U_C$.
$\Gamma[G]$ is extremized by the exact Green's function
and gives the free energy of the system in the extremum.

First, we discuss the Hartree-Fock approximation. In this case,
$\Phi[G]=
\Phi^H[\rho]+\Phi^X[\rho]$, where
$\Phi^H[\rho]=\frac{1}{2}\int_{\vr\vr'} \rho(\mathbf{r})
U_C(\vr-\vr')\rho(\mathbf{r}')$ and
$\Phi^X[\rho]=-\frac{1}{2}\sum_\sigma
\int_{\vr\vr'}\rho^\sigma(\vr,\vr')U_C(\vr-\vr')\rho^\sigma(\vr',\vr)$.
Here $U_C=\frac{2}{|\vr-\vr'|}$ is the
Coulomb repulsion, 
$\rho(\mathbf{r},\mathbf{r}') =T \sum_{n}G(\mathbf{r},\mathbf{r}';i\omega_{n}) e^{i\omega_{n}0^{+}}$
is the single particle density matrix and $\rho(\vr)=\rho(\vr,\vr)$ is the
electronic density.

Next we discuss the LDA approximation, in which $\Phi[G]$ is approximated by the sum of the
Hartree $\Phi^{H}[\rho]$, local exchange
$\Phi^{LDA,X}[\rho]=\int_{\vr} \rho(\vr) \varepsilon_x(\rho(\vr))$
and local correlation $\Phi^{LDA,C}[\rho]=\int_{\vr} \rho(\vr)
\varepsilon_c(\rho(\vr))$ functional.
Here, we use the Slater exchange functional, 
$\varepsilon_x(\rho)=-\frac{3}{2}(\frac{3}{\pi}\rho)^{\frac{1}{3}}$,
and the Vosko-Wilk-Nusair (VWN) parametrization \cite{corr_vw} for the correlation functional $\varepsilon_c(\rho)$. 

DMFT approximates the exact Luttinger
functional $\Phi[G]$ by its local counterpart $\Phi^{DMFT}=\sum_i
\Phi[\cG^i_{local}]$~\cite{kotliar2006rev}, which contains all skeleton Feynman diagrams
constructed from the local Green's function $\cG^i_{local}$ centered on
the local site $i$, and the local Coulomb repulsion 
$U^i_{local}=\bra{\chi_i\chi_i}U_C(\vr-\vr')\ket{\chi_i\chi_i}$.
Notice that the exact functional $\Phi$ and the DMFT
functional $\Phi^{DMFT}$ have exactly the same topological structure 
in terms of Feynman diagrams. The only difference is that $\Phi$ is a functional of full Green's function
$G$ and $U_{C}$ while $\Phi^{DMFT}$ is a functional 
of $\cG^{i}_{local}$ and $U^i_{local}$.
The essential DMFT variable is $\cG^{i}_{local}$, 
which is computed by the projection discussed above, and is being matched with the auxiliary impurity
Green's function.

The flexibility of the DMFT
approximation allows one to treat some parts of the
functional exactly, such as the Hartree-Fock terms.
This approximation is denoted by
HF+DMFT, i.e.,
$\Phi^{HF+DMFT}[G]=
\Phi^{H}[\rho]+\Phi^{X}[\rho]+\sum_i\Phi^{DMFT}[\cG_{local}^i]-\sum_i \Phi^{DC}[\rho^i_{local}]$.
Since local part of the Hartree and exchange term is present also in
DMFT, we have to subtract terms counted twice
$\Phi^{DC}[\rho^i_{local}]=\Phi^{H}[\rho^i_{local}]+\Phi^{X}[\rho^i_{local}]$,
where
$\rho^{i}_{local} = \ket{\chi_i} \bra{\chi_i} \rho \ket{\chi_i}\bra{\chi_i}$, 
$\Phi^{H}[\rho^i_{local}]=\frac{1}{2}\int_{\vr\vr'}
\rho^i_{local}(\vr)U_C(\vr-\vr')\rho^i_{local}(\vr')$ and
$\Phi^{X}[\rho^i_{local}]=-\frac{1}{2}\sum_\sigma \int_{\vr\vr'}
\rho^{\sigma,i}_{local}(\vr,\vr')U_C(\vr-\vr')\rho^{\sigma,i}_{local}(\vr',\vr)$.

Finally, we define the functional of the combination of LDA and DMFT, i.e., LDA+DMFT, in which the DMFT
correlations (truncated to small subset of important degrees of freedom)
and LDA static correlations complement each
other. The functional is
$\Phi^{LDA+DMFT}[G]=\Phi^{H}[\rho]+\Phi^{X}[\rho]+\Phi^{LDA,C}[\rho]+
\sum_i \Phi^{DMFT}[\cG^i_{local}]-\sum_i \Phi^{DC}[\rho^i_{local}]$
 where the exact-exchange functional $\Phi^{X}[\rho]$ is used because 
the non-local exchange is large in molecular systems. The doubly counted correlation term is contained in 
$\Phi^{DC}[\rho^i_{local}]=
\Phi^{H}[\rho^i_{local}]+\Phi^{X}[\rho^i_{local}] +
\Phi^{LDA,C}[\rho^i_{local}]$, where $\Phi^{H}$ and $\Phi^{X}$ are 
defined above, and the double-counted correlation is
\begin{eqnarray}
\Phi^{LDA,C}[\rho^i_{local}]=\int_\vr \varepsilon_c(\rho^i_{local}(\vr))\rho^i_{local}(\vr).
\label{DCF}
\end{eqnarray}
This is the exact intersection between LDA and DMFT approximation
since it parallels the derivation of the DMFT approximation starting
from the exact functional (For details, see Appendix). The double-counting term is hence a ``DMFT"-like approximation to
the LDA correlation functional. Namely, just as the replacement of the total
$G$ by its local counterpart $\cG^i_{local}$ in the exact functional
leads to the DMFT approximation, replacing total $\rho$ by
$\rho^i_{local}$ in LDA functional gives the intersection of the two
methods.

Although LDA+DMFT is very often used in solid-state electronic
structure calculations, such an exact double-counting term has not been
proposed before. This is because in the solids, there is additional
complexity of screening, whereby the core, semicore and other states
excluded by the DMFT model screen the Coulomb interaction in the DMFT correlated space.
Therefore the derivation of the DMFT
Luttinger-Ward functional in solids requires not 
only the substitution of the total Green's
function $G$ by $\cG_{local}$ but also unscreened Coulomb repulsion 
$U_C=\frac{2}{|\vr-\vr'|}$  by screened one
$\cU_C=\frac{2e^{-\lambda|\vr-\vr'|}}{|\vr-\vr'|}$. 

For solid state calculations, we propose to 
approximate the double-counting functional
by a similar functional as in Eq.~\ref{DCF}, but with
$\varepsilon_{c}$ being a function of density and screening length
$1/\lambda$, i.e.,
$\varepsilon_{c}(\rho_{local}(\vr),\cU(\lambda))$. This requires
calculation of the correlation density for the electron gas model by
quantum Monte Carlo in two dimensional space, as a function of density
$r_s$ and screening $\lambda$.  The same screened form of the Coulomb
repulsion has to be then used in the DMFT impurity calculation, i.e.,
$\cU_{m_1 m_2 m_3 m_4} = \bra{m_1
  m_2}\frac{2e^{-\lambda|\vr-\vr'|}}{|\vr-\vr'}\ket{m_3 m_4}$.
Notice that the screening length $\lambda$ is uniquely determined by the screened Coulomb parameter $\cU$, which can be estimated by constrained LDA \cite{clda} or constrained RPA \cite{Aryasetiawan2004}.

\begin{figure}[!t]
\centering{
\includegraphics[width=0.9\linewidth,clip=]{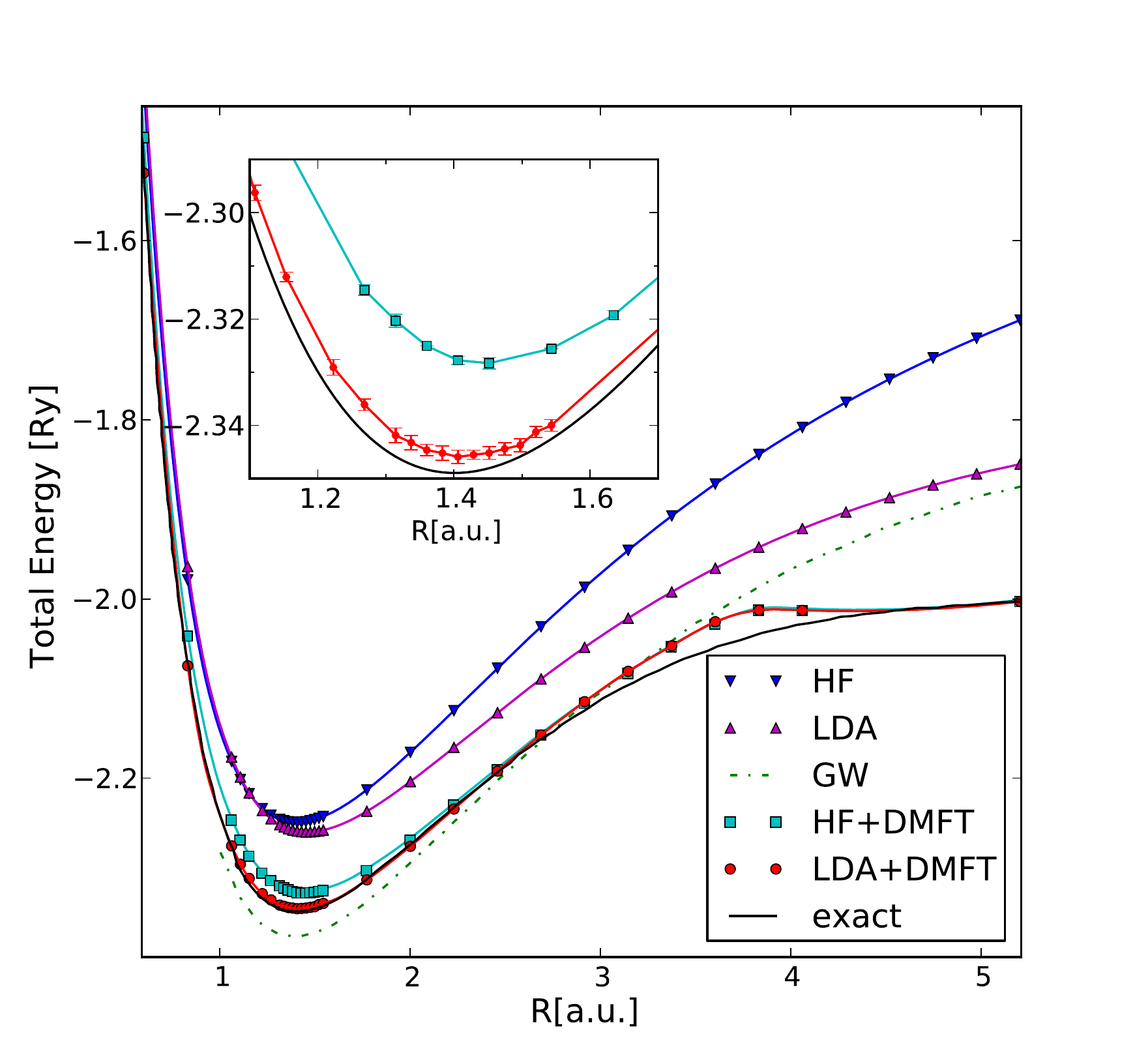}
}
\caption{ (Color online) Total energy curves of the H$_2$ molecule
  versus interatomic distance $R$ calculated by different methods:
  LDA+DMFT, HF+DMFT, LDA, and HF. The GW and
  exact result are also presented for comparison.
}
\label{fig1}
\end{figure}

\section{Results}

In Fig.~\ref{fig1} we compare the total energy curves of the H$_2$ molecule
versus nuclear separation obtained by different electronic structure
methods to the exact result from Ref.~\onlinecite{kolos1964}. The Hartree-Fock method describes the equilibrium distance
quite well ($R\approx 1.39$ compared to exact $1.402$), but the energy
is severely overestimated, in particularly upon dissociation.  This
well-known problem is attributed to static correlation that arises in
situations with degeneracy or near-degeneracy, as in many transition
metal solids and strongly correlated systems.

Due to missing correlations, at large distance the Hartree-Fock method
predicts that the two electrons is both found at one nucleus
with the same probability as finding them away from each at its own
nucleus.
By including static local correlations, the LDA method improves on
the energy at large distance, although it is still way higher than the
energy of two hydrogen atoms.
The equilibrium distance is overestimated
by LDA ($R\approx 1.46$) and the total energy at equilibrium is similar
to its Hartree-Fock value. We also include in the plot the result of
the self-consistent GW calculation from Ref.~\onlinecite{partialGW2009}, which gives
a quite lower total energy at equilibrium and severely
overestimated dissociation energy which is comparable to that of LDA.

At large interatomic separation, the static correlations are not
adequate because of the near-degeneracy of many body states, which can
not be well described by electron density alone. The DMFT uses the
dynamical concept of the Green's function and captures correctly the
atomic limit.  This is because at
large interatomic distance the impurity hybridization function, which
describes the hoping between the two ions, vanishes, and consequently
the impurity solver recovers the exact atomic limit. The inclusion of
dynamic correlations by DMFT (HF+DMFT) also substantially improves the
total energy for all distances, including at equilibrium, and the
error of the total energy is below 1\% for almost all distance, except
around $R\approx 3.6$, where error increases to 2\%. This transition
region close to dissociation is notoriously difficult, because
correlations beyond single site have significant contribution, and
therefore the cluster DMFT is needed to avoid this
error~\cite{millis2011chem}. The predicted equilibrium distance is
slightly overestimated ($R\approx 1.44$).

Finally, the combination of LDA and DMFT gives surprisingly precise
total energy curve. Except around the transition to dissociation ($R\approx
3.6$), it predicts total energy within 0.2\% of the exact result, and
correct equilibrium distance $R \approx 1.4$. Such success of LDA+DMFT 
implies that LDA and DMFT capture complementary
parts of correlations. While DMFT includes all local dynamical correlations
at a single H-ion, it neglects Coulomb repulsion between
electrons that are located at different ions, and poorly describes the
correlations in the regions close to the midpoint, where
$\rho^R_{local}(\vr)$ and $\rho^L_{local}(\vr)$ are comparable in
size. In this case, DMFT correlations are approximated by a linear sum
of two independent terms, the left and right correlations, which
misses the essential non-linearity of the electronic
correlations. This situation is very common in solid state
calculation, where charges solely from the most localized orbitals (such
as $d$ or $f$) are treated by DMFT, while majority of the electronic
charge is described by the LDA correlations. On the other hand,
LDA adds correlations due to all electronic charge, which is a static
and purely local approximation. The two methods are clearly
complementary, and lead to extremely precise total energy when correctly combined.

\begin{figure}[!t]
\centering{
\includegraphics[width=0.8\linewidth,clip=]{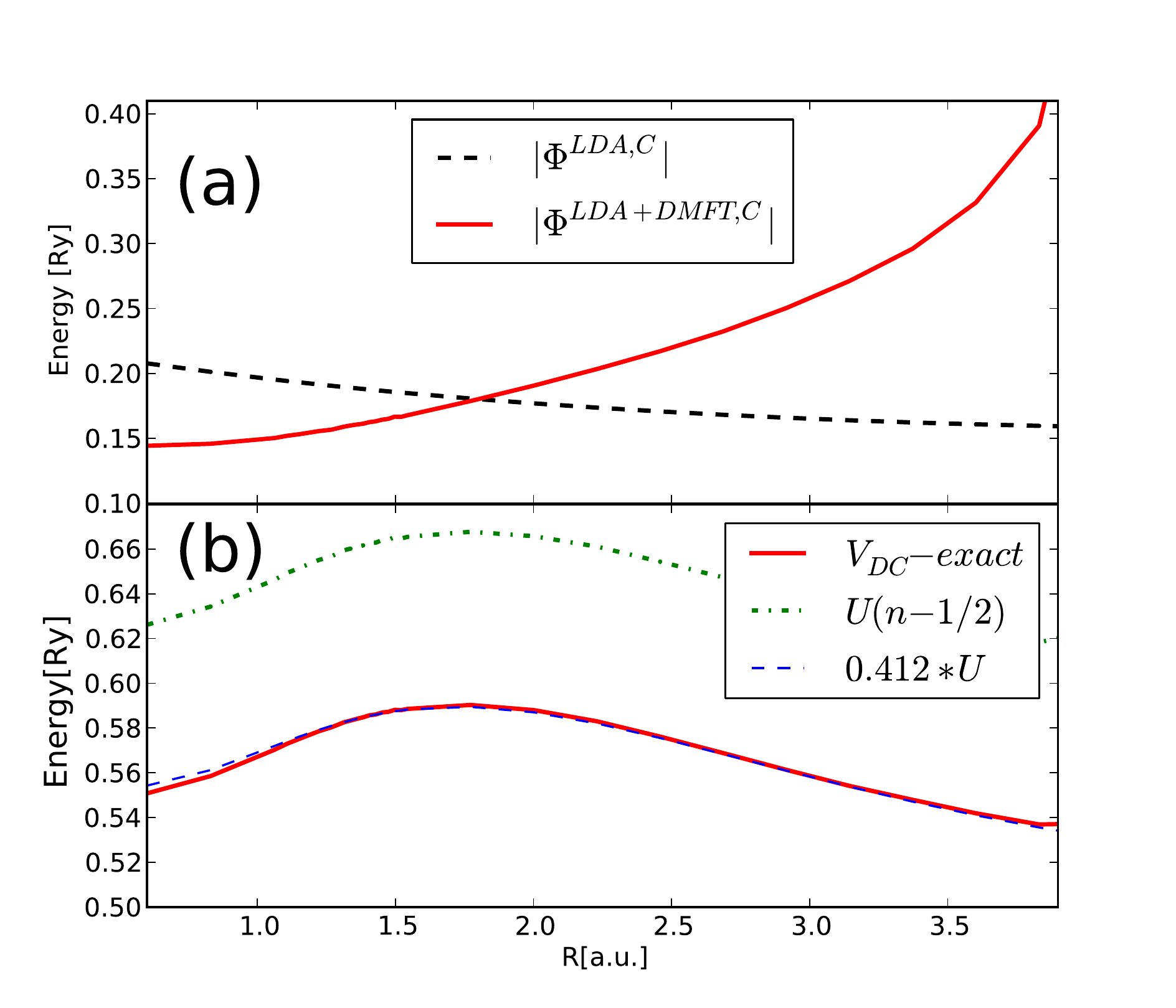}
}
\caption{ (Color online) (a) Correlation energy of LDA and LDA+DMFT
  versus interatomic separation in H$_2$ molecule. The DMFT
  correlation is evaluated by $\Phi^{\mathrm{LDA+DMFT},C}=\Phi^{\mathrm{LDA},C}+\sum_i
  E^{\mathrm{DMFT},i}-\sum_i \Phi^{DC}[\rho^i_{local}]$, where
  potential energy $E^{\mathrm{DMFT},i}=\frac{1}{2}\Tr(\Sigma^{i}_{loc}  \cG^{i}_{loc})$.
  (b) LDA+DMFT double-counting potential $V_{DC}$ versus $R$, which is
  defined as the functional derivative of
  $\Phi^{DC}[\rho^i_{local}]$ given in Eq.~\ref{DCF}.
}
\label{fig2}
\end{figure}

\begin{figure}[h!]
\centering{
\includegraphics[width=1.0\linewidth,clip=]{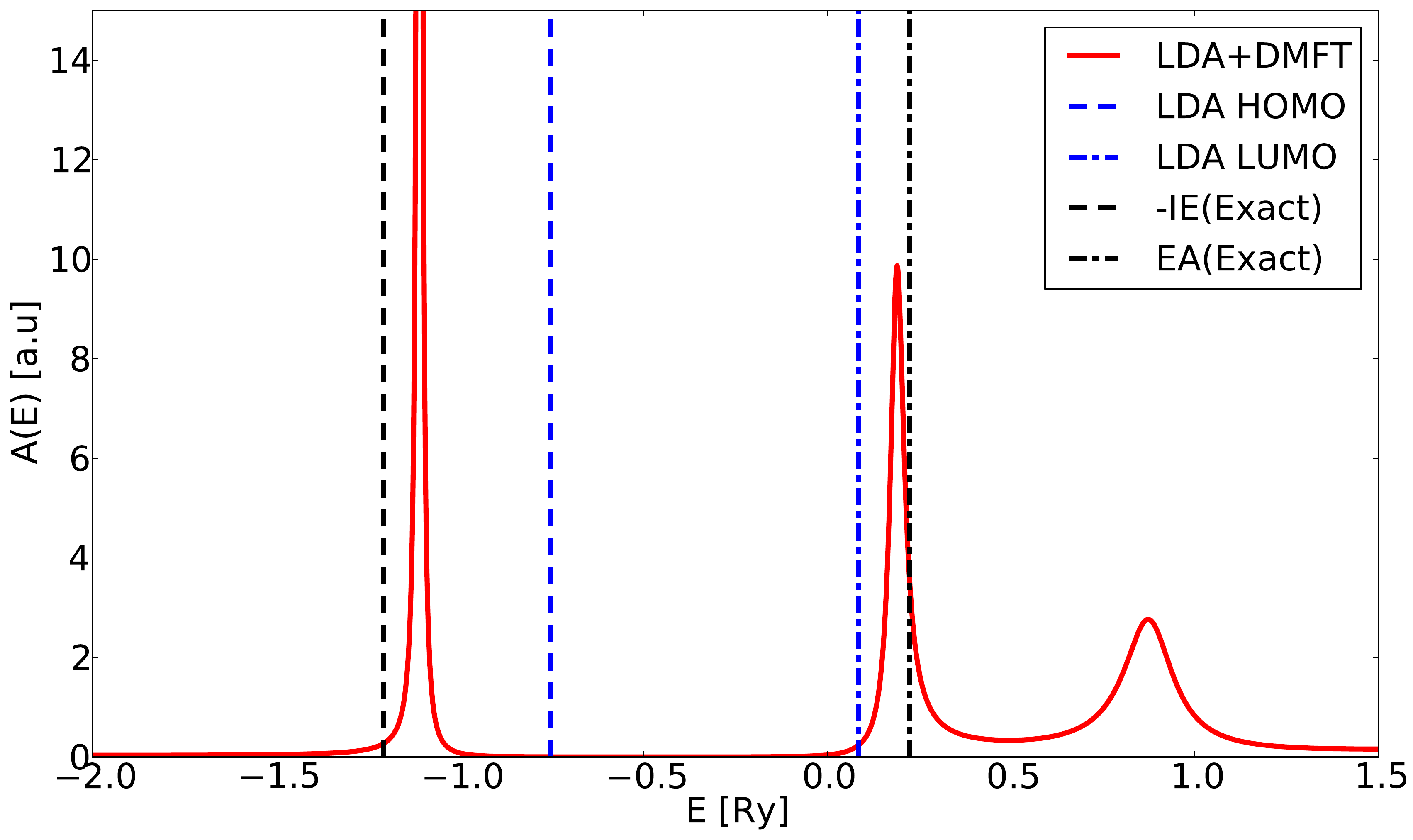}
}
\caption{ (Color online) LDA+DMFT spectral function (red) 
presented with the HOMO and LUMO energy of LDA (blue), and the exact $-$IE and EA (black).}
\label{fig3}
\end{figure}

To gain deeper insight into correlation energy, we
plot in Fig.~\ref{fig2} the correlated energy of the LDA and LDA+DMFT 
versus ion separation. The LDA correlation energy
slightly decreases with increasing distance~\cite{H2XX} in contrast
to physical expectations. On the other hand, the LDA+DMFT correlation is
small when the two ions are close, and it increases sharply with increasing
distance, signaling a Mott-like transition, where we find the DMFT 
self-energy develops a non-analytic pole
between highest occupied molecular orbital (HOMO) and lowest
unoccupied molecular orbital (LUMO).

In the lower panel of Fig.~\ref{fig2}, 
the exact double-counting potential within LDA+DMFT defined as
$V_{DC}=\bra{\chi_i}\delta\Phi^{DC}/\delta\rho^i_{local}\ket{\chi_i}$
versus $R$ is displayed.
The often used phenomenological form $U(n-1/2)$, first introduced in the context of
LDA+U~\cite{dc_fll}, is also shown for comparison. The exact
double-counting is kept somewhat smaller than the phenomenological
form, and its variation is almost entirely due to variation of local
Coulomb repulsion
$U=\bra{\chi_i\chi_i}U_C(\vr-\vr')\ket{\chi_i\chi_i}$, with
proportionality constant $V_{DC}\approx 0.412 U$. In the solid state
calculations, the self-consistent form of the double-counting
$U(n-1/2)$ is also often found too large and is many times reduced (see
discussion in Ref.~\onlinecite{dc_haule}.)

In Fig.~\ref{fig3}, we plot the LDA+DMFT spectral function at equilibrium distance
R = 1.4, which has been analytically continued to the
real axis by Pade method. The highest occupied 
quasi-particle peak (HOMO, below 0) has the physical meaning of 
minus the ionization energy ($-\mathrm{IE} = E_{H_2}-E_{H_2^{+}}$) while the 
lowest unoccupied peak (LUMO, above 0) corresponds to electron affinity 
($\mathrm{EA} = E_{H_2^{-}}-E_{H_2} $).
These quantities are called vertical IE and vertical EA, respectively, 
because these energies of removal/addition of an electron 
are calculated with fixed interatomic separation R.
The exact $-$IE ($-1.207$ Ry) and EA (0.224 Ry) 
are presented as black vertical lines, 
calculated from the total energy difference using
the exact methods  \cite{hadinger1989, kolos1964, anion_fci}.
We also mark the position of LDA HOMO ($-0.754$ Ry)
and LDA LUMO (0.084 Ry) with blue lines.

The LDA HOMO is almost 40\% off the $-\mathrm{IE}$ and the 
LDA LUMO is around 60\% off the EA.
This failure of Kohn-Sham (KS) eigenvalues is due to 
delocalization error of LDA functional, connected with
the well known underestimation of band-gaps by LDA
\cite{homolumo2003,homolumo2007}. On the other hand, the spectral 
function of LDA+DMFT, in which the KS eigenvalues 
are renormalized by DMFT self-energy,
shows a sharp resonance around $-1.110$ Ry (7\% of error bar), a substantial improvement
from 40\% error bar of LDA HOMO. The LUMO peak
is also refined from $0.084 $ Ry (LDA) to $0.192$ Ry which is only 0.032 Ry off the exact value.

Although LDA+DMFT spectral function considerably improves the LDA excitations, 
it still deviates from the exact result (for $-$IE, it is about 7\% off).
In order to obtain an insight into this mismatch of the LDA+DMFT spectral function
and the exact result,
we investigated H$_2$ in two different ways using GW-RPA approach: (a) one 
 considering GW
correlation in the entire Hilbert space and (b) the other where GW
is solely confined to the DMFT projected space (Eq \eqref{loc_orbital}).
Firstly, we found no significant total energy difference ($\sim 0.005Ry$) between two schemes.
On the other hand, for spectral function, the scheme (a) yields its $-$IE peak very close 
to the exact value within 0.1\% error while that of the scheme (b), in which GW correlation is restricted to the minimal DMFT orbitals,
is also 7\% off the exact $-$IE peak.
This indicates that the correlation of the rest of 
the Hilbert space needs to be included to predict an accurate spectra of H$_2$
although the correlation within the minimal orbital set (Eq. \eqref{loc_orbital})
is enough to capture the total energy precisely. We believe that
multi-orbital LDA+DMFT framework, where 
the DMFT correlations are also extended to higher excited states of the system, 
 would lead to progressively better results.

\section{Conclusion}
In summary, a good implementation of LDA+DMFT with a high-quality
projector and exact double-counting that have been introduced here, can rival many
quantum chemistry methods in its precision. In the DMFT case, since the most time
consuming part of the method -- the inclusion of correlations on a
given ion -- scales linearly with the system size, it holds great
promise in future quantum chemistry and solid state applications, although
it still needs to be tested in other molecular systems 
to establish its ultimate usefulness in
quantum chemistry applications.  We have showed that the H$_2$ molecule is
a very good testing ground for electronic structure methods addressing
correlation problem, especially because the screening effects are not
obscuring problems connected with the choice of the functional to be
minimized. The present methodology will be useful in further developing
other electronic structure methods such as GW+DMFT, where the precise
form of the functional, the level of self-consistency, screening, and
double-counting still need to be adequately addressed.


\appendix*
\section{The double-counting functional in detail}

We firstly define the projected local Green's function for the $i$-th atom
using the DMFT projector $\ket{\chi_i}$ (Eq. \eqref{loc_orbital}):

\begin{eqnarray}
G^{i}(\omega) = \bra{\chi_i} G(\omega) \ket{\chi_i}
\end{eqnarray}
where the index $i$ specifies the atomic site $L$ or $R$ and in position space it takes the form of
\begin{eqnarray}
\cG^i(\omega;\vr,\vr')=\chi_i(\vr)  G^i(i\omega)  \chi_i^*(\vr').
\end{eqnarray}
For more complete model, we define projectors containing orbital index $\alpha$
as well as the site index $i$, $\ket{\chi_{\alpha}^i}$,
which is relevant for molecules with heavier atoms.
The local Green's function in DMFT basis is then written as
$G^i_{\alpha\alpha'}= \bra{\chi_{\alpha}^i} G(\omega) \ket{\chi_{\alpha'}^i}$
and its position space version is
$\cG^i(\vr,\vr')=\sum_{\alpha,\alpha'}\chi_{\alpha}^{i}(\vr)G^i_{\alpha\alpha'}\chi_{\alpha'}^{i*}(\vr')$.

The Luttinger-Ward functional for LDA+DMFT approximation is
\begin{align}
\Phi^{LDA+DMFT} &= \\ \nonumber
\sum_i \Phi^{DMFT}&[G^i_{local}] + \Phi^{LDA}[\rho] -
\sum_i \Phi^{DC}[\rho^i_{local}].
\end{align}
Here the local density $\rho^i_{local}$ is defined in the same way as 
$\cG^{i}(\vr,\vr')$ from $G(\vr,\vr')$, namely,
\begin{align}\label{loc_density}
\rho^i_{local}(\vr)&=\chi_i(\vr)  n^i_{local}  \chi_i^*(\vr)
\end{align}
where $n^i_{local}$ is local occupation.

The double-counting energy can be split into Hartree-Fock part $\Phi^{DC}_{HF}$ and the
correlation part $\Phi^{DC}_c$. The Hartree-Fock part is straightforward to
evaluate in DMFT basis as
\begin{align}
\Phi^{DC}_{HF}[\rho^i_{local}] = \frac{1}{2} & \sum_{\sigma\sigma'}
\bra{\chi_i\chi_i}U_C\ket{\chi_i\chi_i}\times \\ \nonumber
&(n^i_{local,\sigma}n^i_{local,\sigma'}-\delta_{\sigma\sigma'}n^i_{local,\sigma}n^i_{local,\sigma}).
\end{align}
When a single orbital per site is considered with the above defined projector, the Hartree-Fock
energy and potential simplify to
\begin{eqnarray}
&& \Phi^{DC}_{HF}[\rho^i_{local}]=\frac{1}{4} U_{local} (n^i_{local})^2\\
&& V^{DC}_{HF}[\rho^i_{local}] \equiv\frac{\delta \Phi^{DC}_{HF}[n^i_{local}]}{\delta n^i_{local}}=\frac{1}{2} U_{local} n^i_{local}
\end{eqnarray}
where $U_{local}=\bra{\chi_i\chi_i}U_C\ket{\chi_i\chi_i}$ is the local Coulomb matrix element.
When multiple orbitals are considered by DMFT, the Hartree-Fock double counting term generalizes to
$\Phi^{DC}_{HF} =
\frac{1}{2}\sum_{\alpha\beta\gamma\delta,\sigma\sigma'}
\bra{\chi_{i\alpha}\chi_{i\beta}}U_C\ket{\chi_{i\gamma}\chi_{i\delta}}( n^\sigma_{\alpha\delta}n^{\sigma'}_{\beta\gamma}-\delta_{\sigma\sigma'}n^\sigma_{\alpha\gamma}n^\sigma_{\beta\delta})$.
where $\alpha\beta\gamma\delta$ run over active orbitals on a given atom.

The double-counting for correlation energy within LDA+DMFT we propose here is
\begin{eqnarray}\label{dc_corr}
\Phi^{DC}_c[\rho^i_{local}]=\int_{\vr} \varepsilon^{LDA}_{c}(\rho^i_{local}(\vr),U_C)\rho^i_{local}(\vr)
\end{eqnarray}
This is exactly DMFT approximation of LDA correlation functional, truncating 
$G(\vr,\vr')  \rightarrow \cG^{i}(\vr,\vr') $ that yields $\rho(\vr) \rightarrow \rho^{i}_{local}(\vr)$.
The expression $\varepsilon^{LDA}_{c}(\rho^i_{local}(\vr),U_C)$ in Eq. \eqref{dc_corr} implies that
it is a functional of both density and Coulomb interaction $U_C(\vr,\vr')=\frac{2}{|\vr-\vr'|}$.
In solid state, we should replace $U_C(\vr,\vr')$ with a screened one $\cU^{\lambda}_C(\vr,\vr')=\frac{2e^{-\lambda|\vr-\vr'|}}{|\vr-\vr'|}$ and therefore we need to obtain 
the LDA correlation functional with respect to two parameters
$\rho$ and $\lambda$ for exact double counting. In small molecular systems such as H$_2$ screening effect is negligible ($\cU^{\lambda}_C(\vr,\vr')\approx U_C(\vr,\vr')$) and therefore 
the functional form of the LDA correlation $\varepsilon^{LDA}_{c}$ in double-counting \eqref{dc_corr} is intact.

The double counting potential $V^i_{DC}$ in the DMFT basis can be easily computed:
\begin{align}
 V^{i}_{DC}&\equiv\frac{\delta \Phi^{DC}_c[\rho^{i}_{local}]}{\delta n^i_{local}} \nonumber \\
 &= \int_{\vr}|\chi_i(\vr)|^2 V_c^{LDA}[\rho^i_{local}(\vr)]
\end{align}
where $V_c^{LDA}$ is the LDA correlation potential that takes the form of $V_c^{LDA} \equiv \varepsilon_c^{LDA}+\delta \varepsilon_c^{LDA} / \delta\rho$. In derivation, we used the relation \eqref{loc_density} and the chain rule. In more general case with multiple local degrees of freedom, the form of local density \eqref{loc_density} should be 
replaced with 
$\rho_{local}(\vr)=\sum_{\alpha,\alpha'}\chi_{\alpha}^{i}(\vr) \rho^i_{\alpha\alpha'}\chi_{\alpha'}^{i*}(\vr)$ where $\rho^i_{\alpha\alpha'}= \bra{\chi_{\alpha}^i} \hat{\rho} \ket{\chi_{\alpha'}^i}$ is projected 
density matrix onto site $i$.

\bibliography{mybib}

\end{document}